\documentclass[preprint,aps,jap,showpacs,showkeys,altaffilletter,frontmatterverbose,floatfix]{revtex4}
\usepackage{amsmath}
\usepackage{amssymb}
\usepackage{graphicx}
\usepackage{dcolumn}
\usepackage{bm}
\usepackage{float}
\usepackage{color}

\DeclareGraphicsExtensions{.pdf,.png,.jpg}
\begin{document}

\preprint{}
\title{ Strain effects in [001] textured Co$_{80}$Ir$_{20}$ thin films with
negative magnetocrystalline anisotropy. }
\author{L. Avil\'{e}s F\'{e}lix, M. V\'{a}squez Mansilla}
\affiliation{Departamento de Magnetismo y Materiales Magn\'{e}ticos (CNEA), Instituto de
Nanociencia y Nanotecnolog\'{\i}a Nodo Bariloche (CNEA - CONICET), Instituto
Balseiro (UNCuyo - CNEA), 8400 Bariloche, R\'{\i}o Negro, Argentina.}
\author{J. E. G\'{o}mez}
\affiliation{Departamento de Magnetismo y Materiales Magn\'{e}ticos (CNEA), Instituto de
Nanociencia y Nanotecnolog\'{\i}a Nodo Bariloche (CNEA - CONICET), 8400
Bariloche, R\'{\i}o Negro, Argentina.}
\author{M. Balod}
\affiliation{Instituto Balseiro (UNCuyo - CNEA), 8400 Bariloche, R\'{\i}o Negro,
Argentina.}
\author{J. Padilla, J. Santiso}
\affiliation{Catalan Institute of Nanoscience and Nanotechnology (ICN2), CSIC and BIST,
Campus UAB, Bellaterra, 08193 Barcelona, Spain}
\author{Subhakanta Das, S.N. Piramanayagam, }
\affiliation{Division of Physics and Applied Physics. NTU, Singapore}
\author{A. Butera}
\email{abutera@cnea.gob.ar}
\affiliation{Departamento de Magnetismo y Materiales Magn\'{e}ticos (CNEA), Instituto de
Nanociencia y Nanotecnolog\'{\i}a Nodo Bariloche (CNEA-CONICET), Instituto
Balseiro (UNCuyo-CNEA), 8400 Bariloche, R\'{\i}o Negro, Argentina.}
\date{\today }

\begin{abstract}
Co$_{80}$Ir$_{20}$ ferromagnetic thin films have recently been the focus of
intensive research because the negative magnetocrystalline anisotropy adds
to the shape anisotropy and favors a strong alignment of the magnetization
in the film plane for [001] textured or epitaxial thin films. However, the
role of magnetoelastic effects has not been properly considered in most
published research. In this work we have performed a detailed analysis of 24
nm Co$_{80}$Ir$_{20}$ thin films deposited on Si/SiO$_{2}$ with different
underlayers (Ta, Pt) and overlayers in order to induce [001]-textured growth
and different degrees of strain. Using x-ray diffraction measurements we
have found that the $c$-axis lattice parameter depends on the underlayer
material (larger negative strain for Ta), but the degree of texture and the
average grain size remain essentially constant, except for one of the
multilayers.

Differences in the magnetic behavior according to the underlayer were also
found in room temperature magnetization vs field loops and temperature
dependent dc magnetization measurements.

Anisotropy was quantified using ferromagnetic resonance which showed that
the effective anisotropy field is also dependent on the underlayer. Ta
underlayers show an anisotropy close to that expected for shape, while Pt
underlayer induces an additional in-plane anisotropy field of the order of
7-9 kOe. A simple model of stress induced anisotropy gives anisotropy field
values similar to those observed experimentally.

The correlation between observed strain and anisotropy together with the
similarity in microstructural properties strongly suggests that stress
effects cannot be disregarded when analyzing the magnetic data for the
estimation of the magnetocrystalline contribution.
\end{abstract}

\maketitle

\section{\label{introduction}Introduction}

\bigskip Ferromagnetic (FM) thin films with negative magnetocrystalline
anisotropy (MCA), which adds to the shape anisotropy and strongly favors the
in-plane alignment of magnetization vector, have been investigated by
different groups\cite{Kikuchi1999,Hashimoto2006,Ngo2011}. The interest in
these systems arises from the potential use in magnetic devices, such as 
microwave applications (soft magnetic films with high resonance frequency
when the magnetic field is applied parallel to the film plane), soft
magnetic underlayers for perpendicular magnetic recording\cite{Ngo2011}, or
the field generator layer in spin torque oscillators\cite%
{Yoshida2010,Nozawa2013}, used for microwave assisted magnetic recording.
Among materials with large negative magnetocrystalline anisotropy Co-rich Co$%
_{100-x}$Ir$_{x}$ with hcp crystalline phase is one of the most studied
systems because it can reach anisotropy values $K_{MC}\sim $ $-7\times 10^{6}
$ erg/cm$^{3}$ for $x\sim 20.$\cite%
{Kikuchi1999,Hashimoto2006,Nozawa2013,Zhang2014} Furthermore, with proper
annealing or high temperature deposition\cite{Nozawa2013} the
magnetocrystalline anisotropy can reach values as large as $K_{MC}\sim $ $%
-9.6\times 10^{6}$ erg/cm$^{3}.$ It was also shown that properly chosen
metallic underlayers are required to promote a [001] fiber texture
perpendicular to the film plane, particularly when using Si substrates with
a native SiO$_{2}$ layer. The preferred underlayers are Ta, Pt and Ru,\cite%
{Hashimoto2006,Ngo2011,Nozawa2013,Wang2013} but Ti and Au have also been
used.\cite{Ma2016,Ma2018} The influence of different seed layers on the
magnitude of $K_{MC}$ was reported\cite{Ma2023} by Ma \textit{et al.} They
use materials with interatomic distances smaller (Ni) and larger (Cu, Ir,
Pt, Au) than that of Co$_{100-x}$Ir$_{x}$ and found a very high [001] fiber
texture in all cases, an almost constant $c-$axis lattice parameter and a
value of $K_{MC}$ that increases when the lattice mismatch was higher. These
observations raise the question whether the observed effective anisotropy is
fully due to shape and magnetocrystalline contributions or if magnetoelastic
effects should be also considered. In order to study the possible effects of
stress on the magnetic anisotropy we present here a study of Co$_{80}$Ir$%
_{20}$ thin films that have been grown on Si/SiO$_{2}$ substrates using
different seed and top layers. We studied the strain effects by means of
x-ray diffraction, rocking curves, $\omega -$scans and Williamson-Hall
plots. For the determination of the general magnetic behavior we used
different magnetometry techniques, while magnetic anisotropy was evaluated
by means of fixed frequency and broadband ferromagnetic resonance (FMR) that
was also used for the determination of $g-$factors, damping parameters and
anisotropy fields.\bigskip \bigskip\ 

\section{\label{ExpDetails}Sample preparation and experimental details.}

Four multilayers with the same Co$_{80}$Ir$_{20}$ (CoIr from now on) nominal
thickness ($t=$ 20 nm) and different Ta and Pt layers were deposited by
magnetron sputtering techniques at room temperature. We used as substrates 
SiO$_{2}$ oxidized (100)\ Si crystals. The stacks structure are:\ CoIr-1:
Si/SiO$_{2}$//Ta(1.5)/CoIr(20)/Pt(5); CoIr-2: Si/SiO$_{2}$%
//Ta(1.5)/Pt(5)/CoIr(20)/Ta(1.5); CoIr-3: Si/SiO$_{2}$%
//Ta(1.5)/CoIr(20)/Ta(1.5); CoIr-4: Si/SiO$_{2}$%
//Ta(1.5)/Pt(5)/CoIr(20)/Pt(5). Number in parentheses indicate the nominal
thickness in nm. Note that for samples CoIr-1 and CoIr-3 the FM is deposited
directly on top of the Ta seed layer while for CoIr-2 and CoIr-4 it is
deposited on top of Pt. On the other hand, films 1 and 4 are finished with a
top Pt layer and films 2 and 3 with a Ta protective layer.

CoIr, Pt and Ta layers were deposited in an AJA\ sputtering system with a
base pressure $<1\times $ 10$^{-8}$ Torr, sputtering power 50 W (power
density 2.5 W/cm$^{2}$), argon pressure 3 mTorr, which resulted in a
sputtering rate of 0.025 nm/s, 0.053 nm/s and 0.012 nm/s, respectively. The
crystallographic structure, lattice parameters, mosaicity, strain, grain
size and thickness of the films was measured by means of x-ray diffraction
(XRD), rocking curves, $\omega -$scans, Williamson-Hall plots and x-ray
reflectivity (XRR) using an X'Pert Pro MRD (Malvern-Panalitycal)
diffractometer.

The magnetic characterization was accomplished by performing dc
magnetization and ferromagnetic resonance experiments. Saturation
magnetization data of CoIr films were measured using a LakeShore model 7300
Vibrating Sample Magnetometer (VSM). Magnetization loops were also measured
in a home made Magneto-Optic Kerr Effect (MOKE) magnetometer. Experiments in
the longitudinal configuration were made using a red laser ($\lambda =632$
nm) with a power of 5 mW. Magnetization loops were measured with $H$
parallel to different in-plane orientations in order to check for the
presence of in-plane anisotropy. Low temperature $M$ vs $T$ data were
collected in a S700 SQUID from Cryogenics at an applied field of 10 kOe.

Ferromagnetic resonance measurements were made either in a Bruker Elexsys
E500 or a Bruker ESP300 spectrometer at X-band (frequency $\sim 9.5$ GHz) at
room temperature. The samples were placed in the center of a rectangular
resonant cavity, where the derivative of the absorbed power was measured
using standard field modulation and lock-in detection techniques with
modulation amplitudes in the range of 20 Oe. The samples could be rotated
inside the resonator in order to collect the spectra for different
orientations of the films with respect to the external magnetic field.
Broadband FMR (1 GHz -\ 20 GHz) was measured using a home made system
consisting of a coplanar waveguide, a SG22000Pro signal generator, a HEROTEK
DZR400KB detecting diode and the magnet system of the Bruker ESP300
spectrometer. The external field was modulated at $\sim $1500 Hz and the
signal detected using a Signal Recovery 7265 lock-in amplifier.

\bigskip

\section{\protect\bigskip Experimental Results}

\subsection{X-ray diffractometry\label{Section XRayDiffractometry}}

\begin{figure}[tbh]
\includegraphics[ width=12cm]{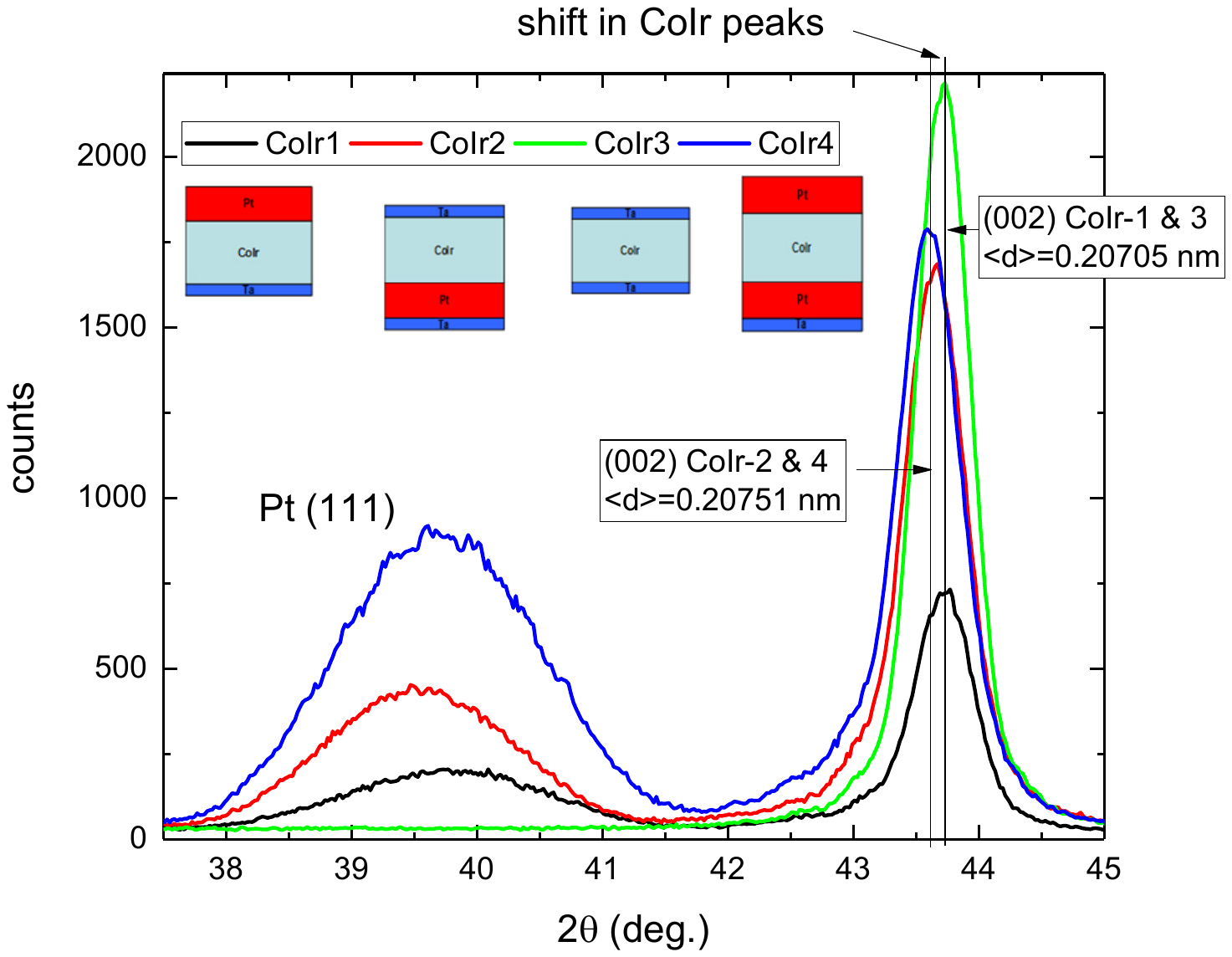}
\caption{X-ray diffractograms in the region where diffractions from CoIr and
Pt are observed. We indicated the observed shift in the samples CoIr-1 and
CoIr-3 compared to CoIr-2 and CoIr-4.}
\label{FigDRX002-111}
\end{figure}

Standard $\theta -2\theta $ diffractograms have been acquired for the four
samples in the angular range $4^{\circ }\leq 2\theta \leq 130^{\circ }$. In
all cases we could only detect reflections associated to CoIr (002) and
(004) planes and Pt (111) and (222), additional to the Si (400) peak coming
from the substrate. In Fig.\ref{FigDRX002-111} we show a selected region of
the diffractograms for the four studied samples. We can see in the figure a
small shift in the angular position of the (002) CoIr-1 and CoIr-3 peaks
compared to those of CoIr-2 and CoIr-4, suggesting the role of the
underlayer in causing a strain. A smaller $d(002)$ interplanar distance was
observed in the case of CoIr-1 and CoIr-3. Considering also the CoIr (004)
reflections we estimated $\left\langle d_{002},2d_{004}\right\rangle
_{1,3}=0.20700(3)$ nm and $\left\langle d_{002},2d_{004}\right\rangle
_{2,4}=0.20752(10)$ nm (see Table \ref{tableI}) from which we can obtain the
relative strain between the two series of films $(d_{1,3}-d_{2,4})/d=-2.5(5)%
\times 10^{-3}.$ The figure also shows that Pt grows with a strong (111)\
texture when it is used both as underlayer or top layer. In the case of Pt
with fcc crystalline structure the (111) is a compact plane that matches the
symmetry of the (002) plane of the hcp CoIr. When Pt is used as underlayer
(films CoIr-2 and CoIr-4) the estimated interatomic distance for the (111)
reflection is 0.262 nm, which is considerably larger than the interatomic
distance in the case of CoIr (0.239 nm). This suggests that a CoIr (002)
film could grow on Pt (111) with an in-plane tensile strain. In the case of
a Ta bottom layer, this element tends to be amorphous for very thin films
which, in principle, should not produce a strain on the CoIr layer. There
are reports\cite{Patel2023} that show a considerable larger angular width of
rocking curves for Co films deposited on 1.5 nm of Ta compared to films
grown on 20 nm of Pt. In our case we have found the full width at half
maximum of the rocking curves reported in Table \ref{tableI}. CoIr films 2
and 4, which are deposited on Pt, show relatively narrow rocking curves
indicating a strong [001] fiber texture. The same happens for CoIr-3,
although it is deposited on Ta. For CoIr-1, which is also deposited on Ta,
the width of the rocking curve is at least twice that of the other films
suggesting a more disordered texture which should reduce the
magnetocrystalline contribution to the magnetic anisotropy. The degree of
texture can be also estimated from the ratio between the areas of the CoIr
(002) and the Si (400)\ peaks, which is similar for films CoIr-2, 3 and 4
but considerable smaller in the sample CoIr-1, again indicating a lower
degree of texture in this film. Note, however, that the CoIr(200)/Pt(111)
intensity ratio remains almost constant in all films, implying that both
layers in the samples have a similar degree of crystalline texture.

\begin{table}[h]
\caption{Parameters of all CoIr samples obtained from XRD, XRR and rocking
curves measurements. The intensity ratio compares the CoIr (002), Pt (111)
and Si (400) reflections.}
\label{tableI}\centering%
\begin{tabular}{ccccccc}
sample & thickness (nm) & $\frac{1}{2}d_{002}+d_{004}$ (nm) & 1000$I_{%
\mathrm{CoIr}}/I_{\mathrm{Si}}$ & $I_{\mathrm{CoIr}}/I_{\mathrm{Pt}}$ & $%
d_{grain}$(nm) & rocking curve width \\ 
CoIr-1 & 26(2) & 0.20699(3) & 26.5 & 1.4 & 20 & 11.4$%
{{}^\circ}%
$ \\ 
CoIr-2 & 23(2) & 0.20742(3) & 70.3 & 1.8 & 19 & 4.9$%
{{}^\circ}%
$ \\ 
CoIr-3 & 25(2) & 0.20701(4) & 62.0 & x & 23 & 5.0$%
{{}^\circ}%
$ \\ 
CoIr-4 & 22(2) & 0.20760(4) & 67.3 & 1.5 & 18 & 4.7$%
{{}^\circ}%
$%
\end{tabular}%
\end{table}
From the FWHM linewidth of the CoIr (002) reflection (subtracting the
instrumental linewidth) and using the Scherrer formula we estimated the
grain sizes presented in Table \ref{tableI}. All samples have similar grain
sizes with an average value $d_{grain}^{\mathrm{CoIr}}=20(2)$ nm, which is
of the order of the film thickness. Pt grain sizes were obtained from the
(111)\ reflection and present also similar grain sizes with an average value 
$d_{grain}^{\mathrm{Pt}}=5.2(2)$ nm, again similar to the film thickness.
Measurements for the estimation of the different thicknesses composing the
multilayers was performed with XRR and GenX 3 software\cite{GenX3} was used
to obtain the individual thicknesses. An average CoIr thickness of 24(2) nm
was obtained, somewhat larger than the nominal 20 nm value. For Pt the
average layer thickness was 4.4(7) nm and for Ta 1.6(9) nm, close to the
nominal values. 
\begin{figure}[tbh]
\includegraphics[ width=12cm]{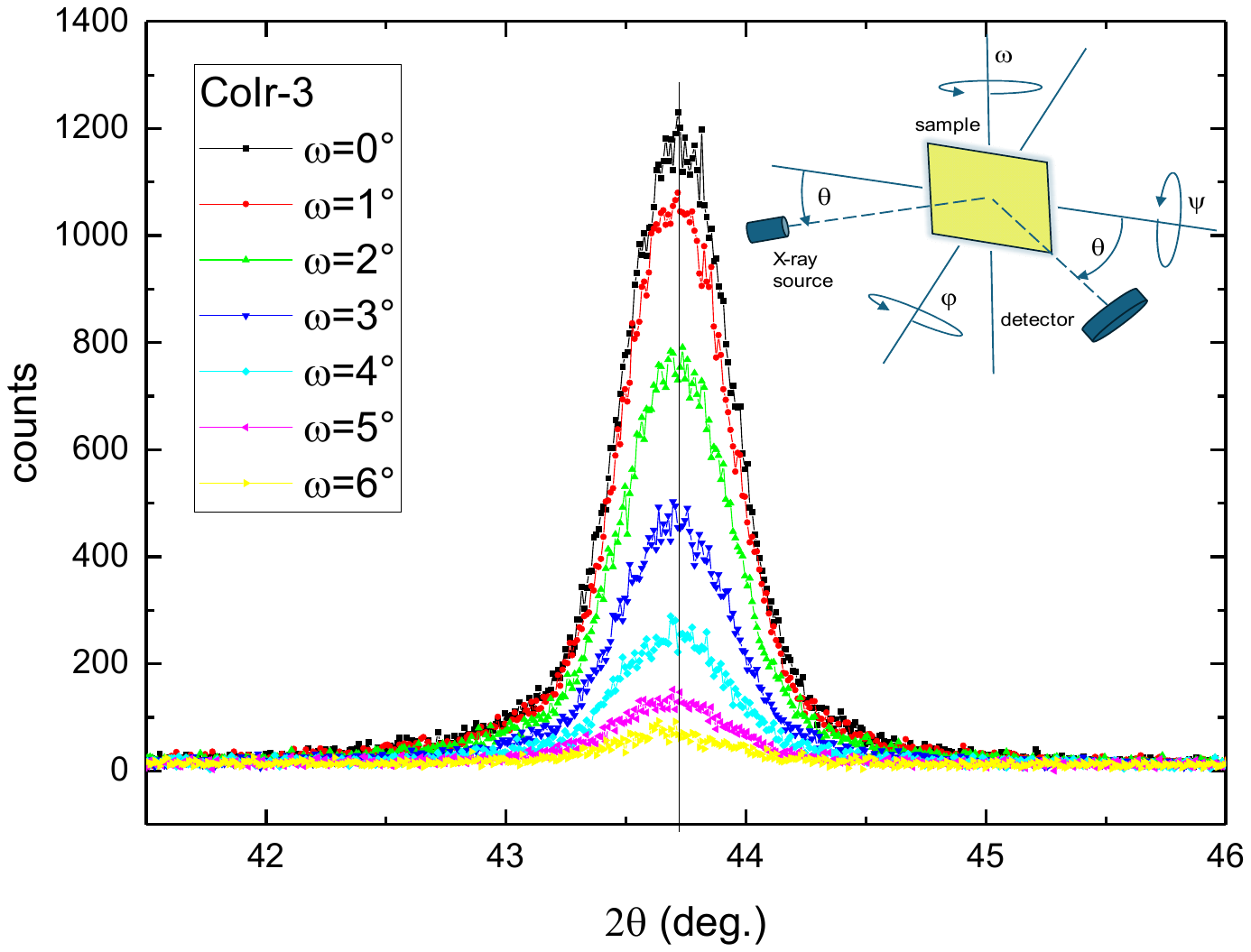}
\caption{CoIr-3 (002) reflection for different values of the angle $\protect%
\omega $ in the range $0^{\circ }\leq \protect\omega \leq 6^{\circ }$. The
vertical line indicates the postion for $\protect\omega =0^{\circ }$. In the
inset we present a schematic diagram of the diffraction set-up}
\label{FigCoIr-3-wscan}
\end{figure}

In order to check for the presence of uniform strain, we performed the
so-called $\sin ^{2}\omega $ (or $\sin ^{2}\psi $) scans\cite%
{Hsiao2009Strain} around the (002) CoIr peak. In this kind of measurements
the sample normal is tilted from the perpendicular alignment in order to
analyze the variation of the interplanar distance for different orientations
of a given set of planes (see inset of Fig. \ref{FigCoIr-3-wscan}). Due to
the strong fiber texture of our films, the peak intensity decreases rapidly
with the angle $\omega $, as can be seen in Fig. \ref{FigCoIr-3-wscan} for
the sample CoIr-3. In all cases we have found a small shift to lower angles
which indicates that the CoIr $d_{002}$ interplanar distance increases for
planes that are not parallel to the film plane. This observation suggests
that all samples can be subjected to a tensile in-plane stress. The
orientation dependence of the interplanar distances can be analyzed using
the $\sin ^{2}\omega $ method which is often used to estimate the uniform
strain in thin films in the out of plane direction ($\varepsilon _{33}$) for
a given $\phi $ and tilt angle $\omega $ (or $\psi )$ which, in the case of
a biaxial strain, can be obtained from the following expression\cite%
{Anderoglu2004,Alvarez2015,Alvarez2016}.

\begin{equation}
\frac{d_{\phi \omega }-d_{0}}{d_{0}}=\left( \varepsilon _{11}\cos ^{2}\phi
+\varepsilon _{12}\cos 2\phi +\varepsilon _{22}\sin ^{2}\phi -\varepsilon
_{33}\right) \sin ^{2}\omega +\varepsilon _{33},  \label{Eq.Strain0}
\end{equation}%
where $\varepsilon _{ij}$ are the strain components and $d_{0}$ is the
relaxed interplanar spacing. The index \textquotedblleft
33\textquotedblright\ indicates a direction normal to the film plane, and
\textquotedblleft 11\textquotedblright\ and \textquotedblleft
22\textquotedblright\ refer to orthogonal axes contained in the film plane.
The $d_{0}$ lattice spacing can be estimated by assuming zero shear stresses
($\varepsilon _{12}=0$) and in-plane isotropic films ($\varepsilon
_{11}=\varepsilon _{22}$), as verified in our case from magnetic
measurements. In the isotropic case the $\phi $ dependence in Eq. \ref%
{Eq.Strain0} disappears to give,%
\begin{equation}
d(\omega )\approx \left( \varepsilon _{11}-\varepsilon _{33}\right)
d_{0}\sin ^{2}\omega +\left( \varepsilon _{33}+1\right) d_{0}\approx -\frac{%
\nu +1}{2\nu }\varepsilon _{33}d_{0}\sin ^{2}\omega +\left( \varepsilon
_{33}+1\right) d_{0}.  \label{Eq.dSin2w}
\end{equation}

In the above equation we assumed an isotropic crystal grain distribution to
relate $\varepsilon _{11}=$ $\frac{\nu -1}{2\nu }\varepsilon _{33}$ through
the Poisson's ratio. As we have already shown, the CoIr layers have a strong
[001] fiber texture which should be accounted for with the corresponding
elastic/compliance constants. In the present case, the stress in the film
normal direction and the shear stresses that involve a component in this
direction are zero.\cite{Clemens1992} With these considerations, the
equation of elasticity in the case of biaxial stress\cite{Anderoglu2004}
takes the following form:%
\begin{equation}
d(\omega )\approx -(1-\frac{S_{11}+S_{12}}{2S_{13}})\varepsilon
_{33}d_{0}\sin ^{2}\omega +\left( \varepsilon _{33}+1\right) d_{0},
\label{Eq.dSin2wS}
\end{equation}%
with $S_{ij}$ the compliance constants of the ferromagnetic film. As far as
we know, there are no reported values of the Poisson%
\'{}%
s ratio or the elastic/compliance constants in Co$_{80}$Ir$_{20}$, but we
can use the accepted\cite{Sander1999,Oliveira2026} values of pure Co ($\nu
=0.32,$ $S_{11}=4.75\times 10^{-3}$ (GPa)$^{-1},$ $S_{12}=-2.33\times 10^{-3}
$ (GPa)$^{-1},$ $S_{13}=-0.69\times 10^{-3}$ (GPa)$^{-1}$) for zero order
estimations, always keeping in mind that CoIr has negative MCA, and that the
change from positive to negative MCA at the spin reorientation transition in
pure Co is accompanied by significant changes\cite{Bidaux1991Coelastic} in
the elastic constants. With these values of pure Co the factor preceding the
term $\varepsilon _{33}d_{0}\sin ^{2}\omega $ in Eqs. (\ref{Eq.dSin2w}) and (%
\ref{Eq.dSin2wS}) equals to 2.06 and 2.75, respectively, stressing the
importance of considering the fiber texture of the films when evaluating
strain effects.

\begin{figure}[tbh]
\includegraphics[ width=12cm]{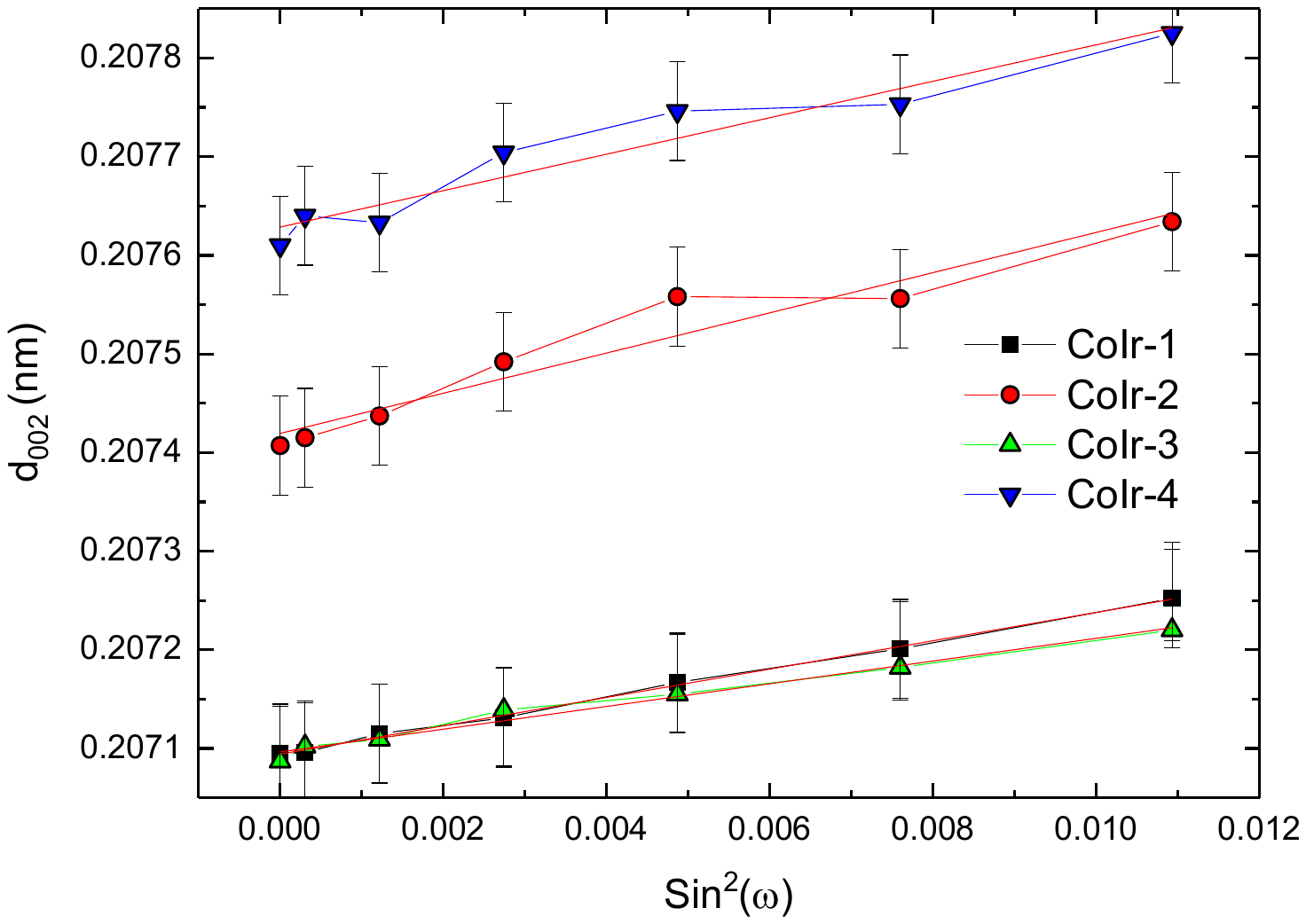}
\caption{Interplanar (002) distance of CoIr peaks for different $\protect%
\omega $ values. The positive slope indicates that $d_{002}$ is minimum for
planes parallel to the film surface while the ordinate can be used to
estimate the strain $\protect\varepsilon _{33}$.}
\label{FigSin2w}
\end{figure}

The evolution of the CoIr $d_{002}$ interplanar distance as a function of $%
\sin ^{2}\omega \ $is presented in Fig. \ref{FigSin2w} where we can observe
an approximate linear behavior with a positive slope in all films,
consistent with a negative value of $\varepsilon _{33}$ and consequently a
tensile in-plane stress. However, the limited angular range in which the
CoIr (002) diffraction can be detected prevented the precise determination
of the slope and hence a consistent estimation of $d_{0}$ and $\varepsilon
_{33}$ from the experimental data. In any case, the positive slope indicates
that $\varepsilon _{33}<0$ and a lower limit for $d_{0}$ is given by the $%
d(\omega =0^{\circ })$ value obtained for the sample CoIr-4 (the film with
the largest $d_{002}$). If, for example, we set $d_{0}=0.2077$ nm we obtain
from the last term of Eq. (\ref{Eq.dSin2wS}) $\varepsilon _{33}(1)\approx
-2.9\times 10^{-3},$ $\varepsilon _{33}(2)\approx -1.3\times 10^{-3}$, $%
\varepsilon _{33}(3)\approx -2.9\times 10^{-3}$, $\varepsilon
_{33}(4)\approx -0.3\times 10^{-3},$ for CoIr-1 to CoIr-4, respectively.
Larger or smaller strain values can be obtained if different values of $d_{0}
$ are used, but in all cases $\varepsilon _{33}<0$ and a larger absolute
value is obtained for samples CoIr-1 and 3. 

\begin{figure}[tbh]
\includegraphics[ width=12cm]{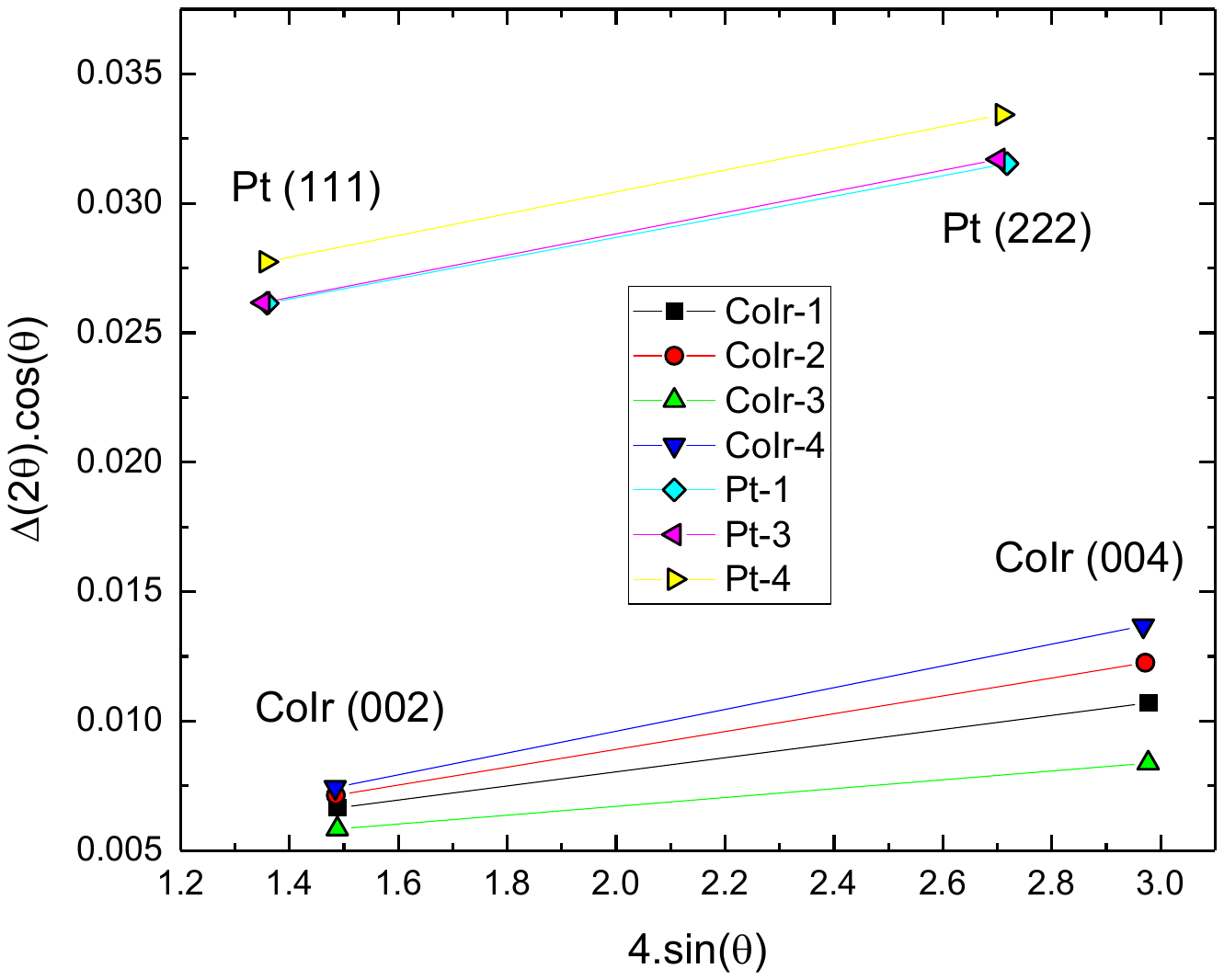}
\caption{Full width at half maximum diffraction linewidth as a function of $%
\sin \protect\theta .$ Grain size and microstrain for the different layers
of the film can be obtained from the CoIr (002) and (004) and Pt (111) and
(222) reflections.}
\label{FigW-H}
\end{figure}

We have also made a Williamson-Hall analysis (see Fig. \ref{FigW-H}) to
separate grain size from microstrain contributions to the diffraction
linewidth. We used the CoIr (002) and (004) and Pt (111) and (222) peaks and
fit the linewidth as a function of the diffraction angle using the
expression: 
\begin{equation}
\Delta (2\theta )\cos \theta =\frac{K\lambda _{Cu}}{d_{grain}}+4\sigma
_{\varepsilon }\sin \theta ,  \label{Eq.W-H}
\end{equation}%
with $\Delta (2\theta )$ the FWHM of the diffraction peaks, $\lambda _{Cu}$
the Cu-K$_{\alpha }$ wavelength, $K\approx 0.9$ (shape factor for thin
films) and $\sigma _{\varepsilon }=\sigma _{d}/d$ the microstrain related to
the line broadening of diffraction peaks. For the Pt layer we obtained
similar values for the grain sizes and the microstrain in all samples with
an average $\left\langle d_{grain}^{\mathrm{Pt}}\right\rangle =6.6(2)$ nm,
slightly larger than the value estimated by using only the (111) diffraction
peak. The average value of microstrain was $\left\langle \sigma
_{\varepsilon }^{\mathrm{Pt}}\right\rangle =4.1(2)\times 10^{-3}.$ In the
case of CoIr, grain sizes were larger than those estimated using only the
(002) diffraction linewidth with values $\left\langle d_{grain}^{\mathrm{CoIr%
}}\right\rangle =36(8)$ nm and an average microstrain $\left\langle \sigma
_{\varepsilon }^{\mathrm{CoIr}}\right\rangle =3(1)\times 10^{-3}$. Note that
the estimated microstrain is comparable in both layers composing the film
and corrects only marginally the grain size in the case of Pt, which
presents relatively wide peaks. For CoIr the contribution of microstrain to
the linewidth is larger and hence a larger grain size is estimated. 

The above presented results strongly suggests that the CoIr/Pt growing order
does not affect significantly grain sizes or microstrain, but uniform strain
is quite different in samples CoIr-1 and 3 compared to CoIr-2 and 4. In all
cases a negative out of plane strain was estimated for the CoIr films. Also,
CoIr-1 has a much broader rocking curve than the rest of the films
indicating a lower degree of [001] fiber texture.

\subsection{DC magnetization measurements}

\begin{figure}[tbh]
\includegraphics[ width=12cm]{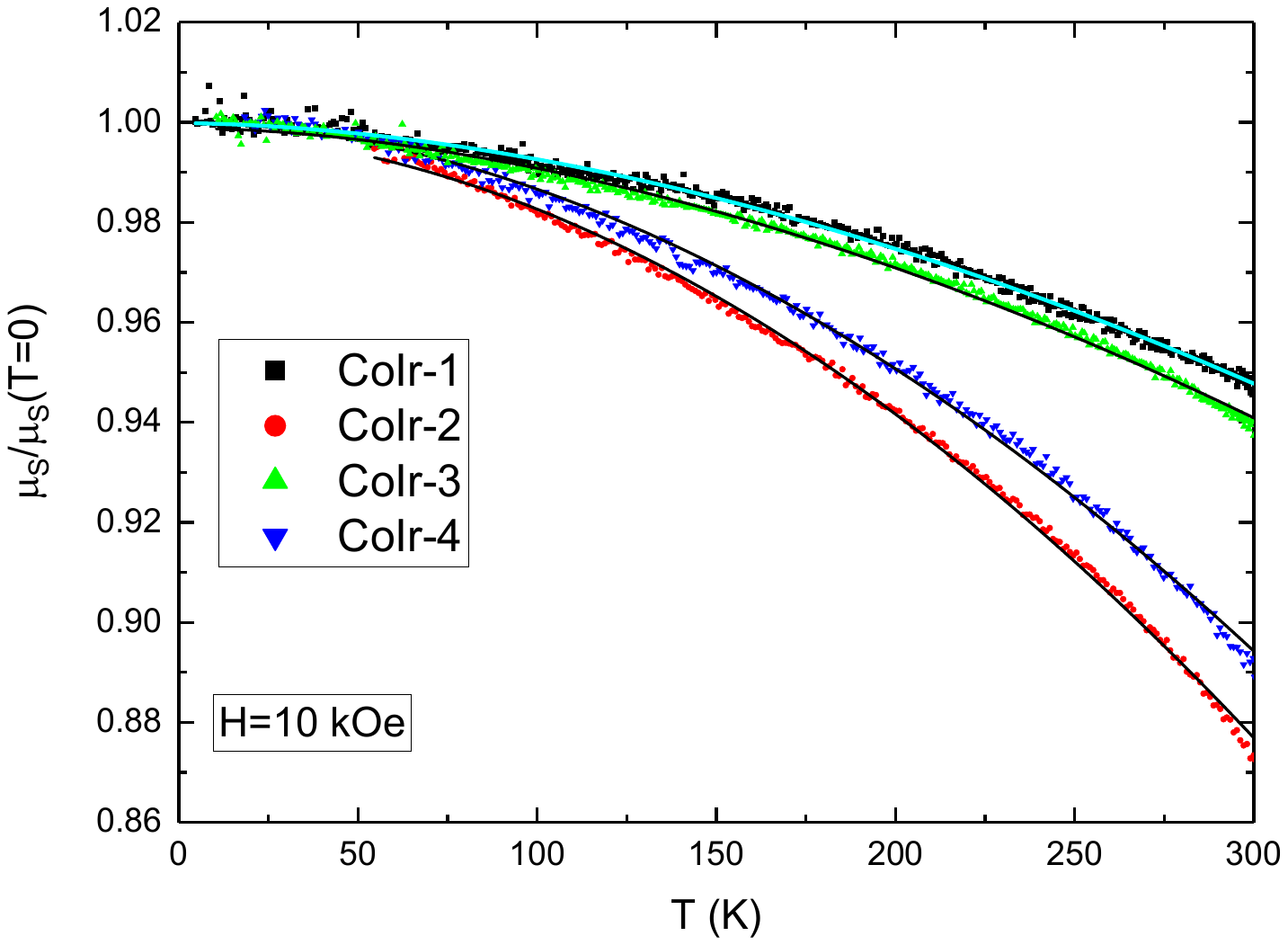}
\caption{Temperature variation of the normalized magnetic moment of all
samples measured at 10 kOe. Continuous lines are the best fits of the
experimental data using Eq. \protect\ref{Eq.MuS(T)}}
\label{FigMvsT1TNorm4K1}
\end{figure}

We show in Fig. \ref{FigMvsT1TNorm4K1} the dependence of the normalized
saturation magnetic moment, $\mu _{S}(T)/\mu _{S}(T=0$ K), in the
temperature range 4 K - 300 K with an applied magnetic field $H=10$ kOe. We
can see that samples CoIr-1 and 3, and CoIr-2 and 4 present a similar
temperature dependence, with an average normalized magnetic moment $%
\left\langle \mu _{S}(T=300\text{ K})/\mu _{S}(0)\right\rangle _{1,3}=0.945$
and $\left\langle \mu _{S}(T=300\text{ K})/\mu _{S}(0)\right\rangle
_{2,4}=0.885$, respectively. This indicates that a small difference in the
room temperature saturation magnetization ($\approx 6$ \%) should be
expected when comparing both sets of films. In practice, small variations in
the CoIr thickness and/or uncertainties in the determination of the sample
area prevent the observation of these differences and we used for the four
samples the same room temperature average magnetization value $M_{s}=$ $%
910\pm 40$ emu/cm$^{3}$ (obtained from VSM measurements and considering an
average thickness $t=24$ nm, as determined from XRR). This $M_{s}$ value is
similar to those usually reported in thin films ($M_{s}=$ $920-980$ emu/cm$%
^{3}$),\cite{Hashimoto2006,Wang2013} the difference probably arising due
small variations in the alloy composition and/or to the formation of an
interfacial dead layer at the film surfaces. Considering an empirical power
law for the temperature dependence of the form%
\begin{equation}
\mu _{S}(T)/\mu _{S}(0)=1-(T/T_{C})^{n},  \label{Eq.MuS(T)}
\end{equation}%
it is possible to reasonably fit the experimental data of all samples. We
obtained similar values of the exponent $n\approx 1.85,$ but significantly
larger $T_{C}$ values for CoIr-1 and 3 ($\left\langle T_{C}\right\rangle
_{1,3}\approx 1500$ K, $\left\langle T_{C}\right\rangle _{2,4}\approx 950$
K). Note that these extrapolated $T_{C}$ values are only indicative of a
different magnetic behavior and are significantly larger than the reported
magnetic order temperature of Co$_{80}$Ir$_{20}$ alloys, $T_{C}\approx 1000$
K.\cite{Massalski198CoIr} The exponent $n=1.85$ is larger than $n=1.5$
expected for the classical $T^{3/2}$ Bloch law due to magnon excitations of
the spontaneous magnetization, but larger $n$ values were found\cite%
{Argyle1963MvsT,Aldred1975MvsT} in metallic ferromagnets and usually
explained by considering the presence of internal or external applied
magnetic fields and an additional $T^{2}$ contribution of conduction
electrons to magnetic excitations. 
\begin{figure}[tbh]
\includegraphics[ width=12cm]{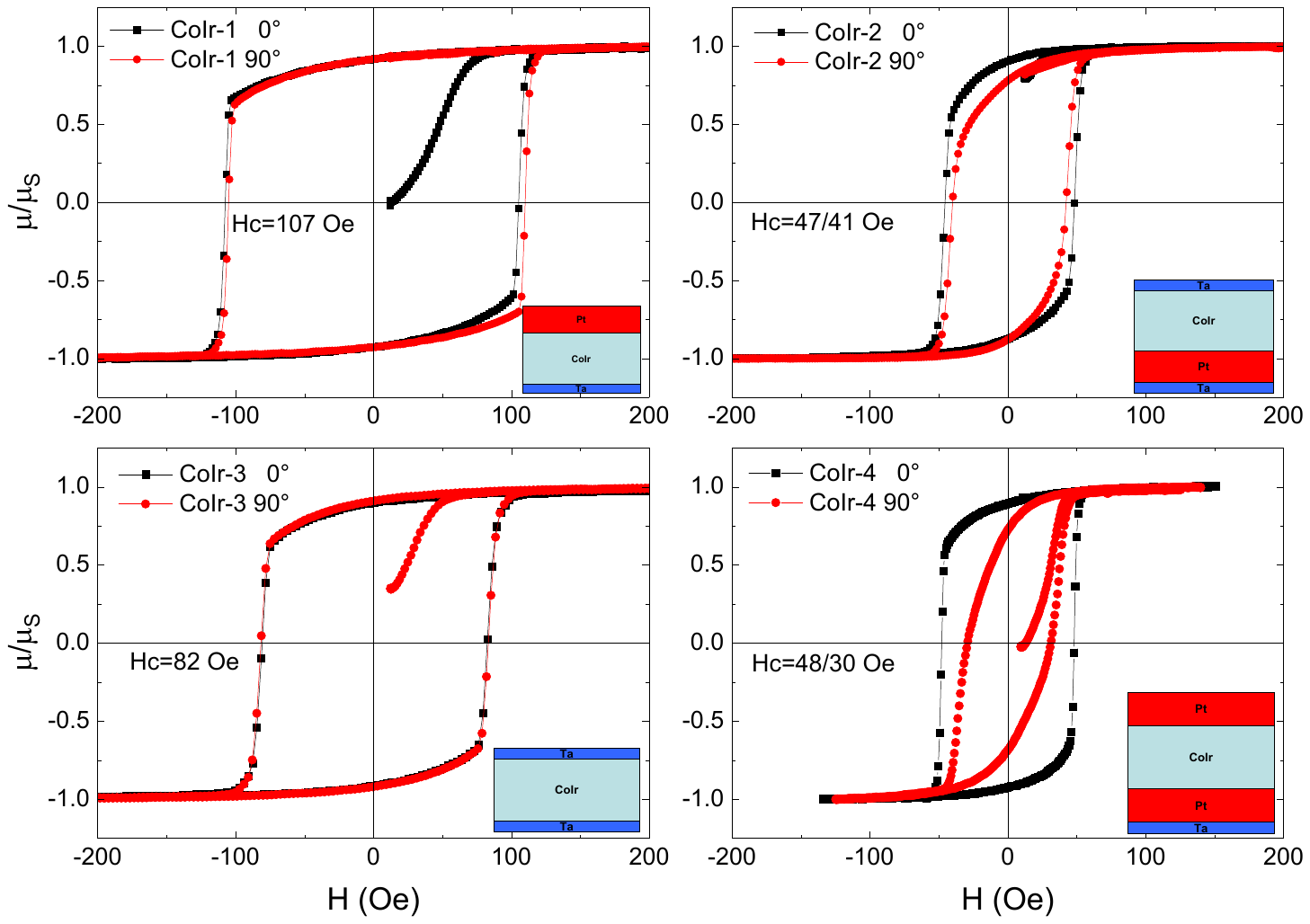}
\caption{Normalized $M$ vs. $H$ loops measured with the field applied
parallel to the film plane in two orthogonal directions. Data was acquired
using longitudinal MOKE magnetometry at room temperature.}
\label{FigMvsH-Kerr}
\end{figure}
We also analyzed all films through room temperature $M$ vs. $H$ loops
measured with MOKE\ magnetometry at different in-plane orientations (see
Fig. \ref{FigMvsH-Kerr}). We can observe that samples CoIr-1 and 3 are
isotropic while a small in-plane uniaxial anisotropy is found in samples
CoIr-2 and 4. Although all films show a relatively high squareness ($%
Sq\approx 0.9),$ CoIr-1 and 3 present an average coercive field that is
twice as large than that of samples CoIr-2 and 4 ($\left\langle
H_{C}\right\rangle _{1,3}\approx 95$ Oe, $\left\langle H_{C}\right\rangle
_{2,4}\approx 47$ Oe). The larger coercivity in CoIr-1 and 3 films
correlates with the larger strain and the higher $T_{C}$ found in these
samples, as already discussed.

\bigskip

\subsection{\protect\bigskip Ferromagnetic Resonance}

\begin{figure}[tbh]
\includegraphics[ width=12cm]{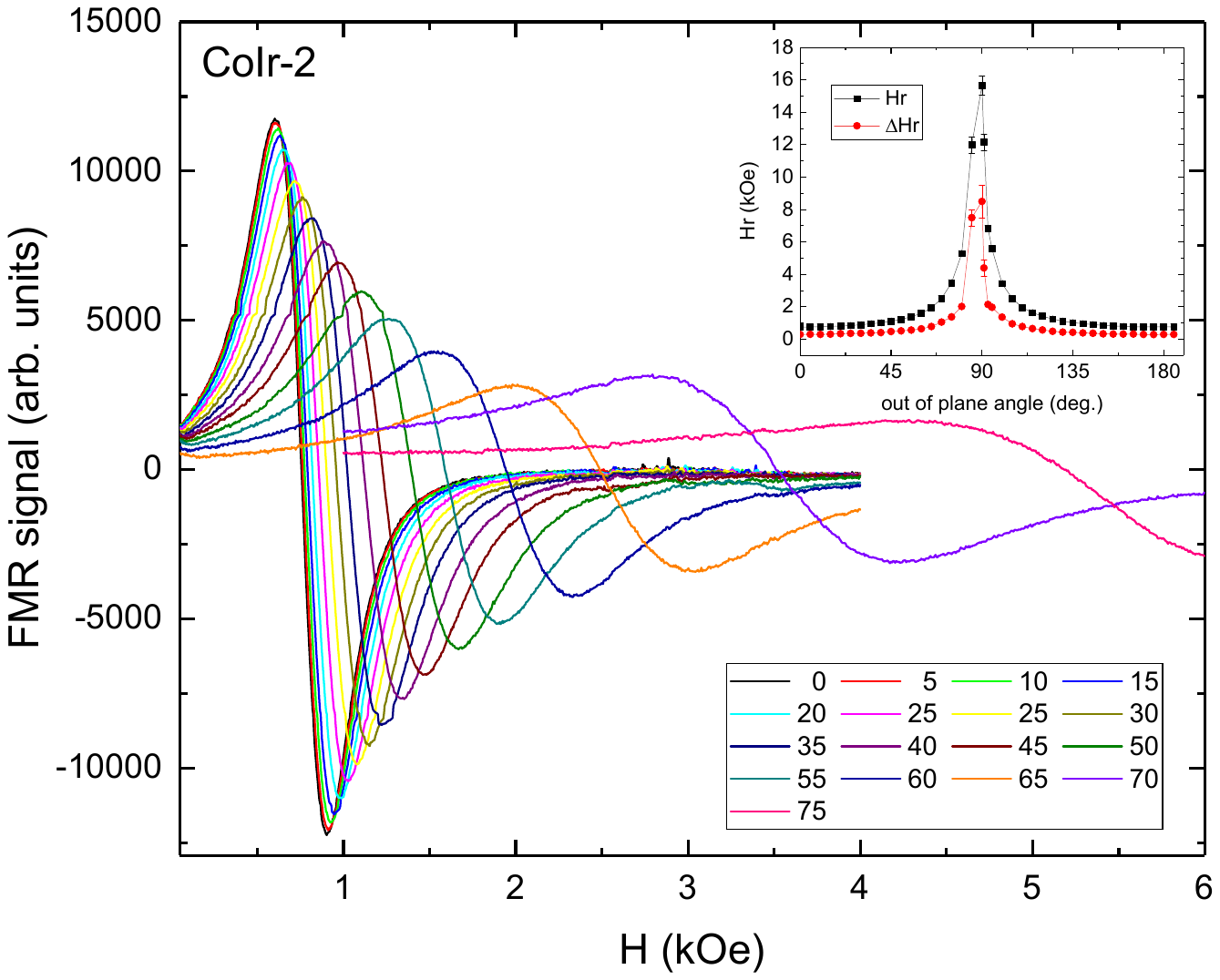}
\caption{Room temperature FMR spectra for different out of plane field
orientations for the sample CoIr-2 measured at X-band. In the inset we show
the angular dependence of the resonance field and the peak to peak line
width.}
\label{FigCoIr-2-spectra}
\end{figure}

\begin{figure}[tbh]
\includegraphics[ width=12cm]{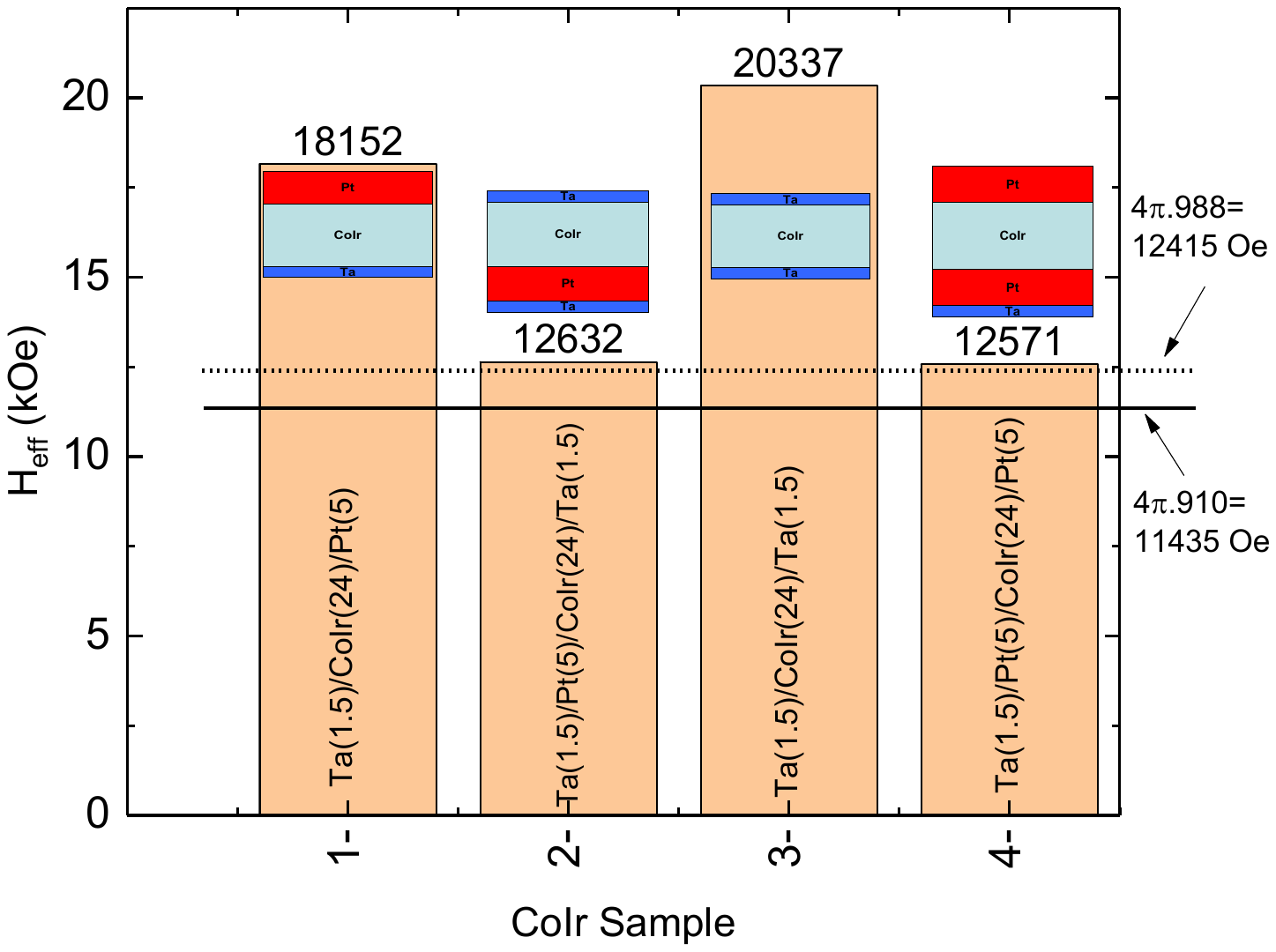}
\caption{Effective anisotropy field obtained from the FMR\ resonance field
in the in-plane configuration. We have used the value of $g=2.2$ deduced
from the angular variation of the sample CoIr-2. The continuous horizontal
line corresponds to the shape anisotropy field calculated with $M_{s}=910$
emu/cm$^{3}$ estimated from our VSM measurements. The dotted line is the
same field obtained using the reported value of $M_{s}$. \protect\cite%
{Hashimoto2006,Wang2013}}
\label{Fig-Heff}
\end{figure}

We have measured all samples in a resonator at X-band $(f\approx 9.5$ GHz)
performing in-plane and out of plane angular variations in order to estimate
the magnetic anisotropy. In all cases we observed a single FMR absorption
line associated to the uniform resonance mode. In-plane angular variations
show an almost isotropic resonance field, coincident with MOKE measurements,
which suggest that there is no (or very small) in-plane crystalline texture
or deposition induced anisotropy. Out of plane angular variations, on the
other hand, show a behavior dominated by the shape anisotropy, with
potential magnetocrystalline and/or stress contributions. In Fig. \ref%
{FigCoIr-2-spectra} we show typical resonance lines measured with $H$ at
different orientations from the film plane in the sample CoIr-2. The
absorption, as expected, moves to larger fields, while the linewidth
increases and the signal amplitude drops significantly as the external field
is rotated to the direction perpendicular to the film plane. In the inset of
Fig. \ref{FigCoIr-2-spectra} we present the full angular variation of $H_{r}$
and $\Delta H_{r}$ for this film. Both magnitudes have a similar angular
dependence with minima in the film plane and maxima in the film normal. This
same behavior was observed in all films. However, for samples CoIr-1 and 3
the maximum available external field (19 kOe) was not large enough to reach
the resonance condition. 
The Smit and Beljers \cite{smit1955FMR,Butera2006FMR} formalism can be used
to obtain the FMR dispersion relation from which it is possible to deduce
the anisotropy fields and the $g-$value using the excitation frequency $%
f=\omega /2\pi $ and the resonance fields measured, at least, along in-plane
and out of plane directions. The dispersion relation can be obtained from 
\begin{equation}
\left( \frac{\omega }{\gamma }\right) ^{2}=\frac{1}{M_{s}^{2}\sin ^{2}\theta 
}\left( \frac{\partial ^{2}E}{\partial \theta ^{2}}\frac{\partial ^{2}E}{%
\partial \varphi ^{2}}-\frac{\partial ^{2}E}{\partial \theta \partial
\varphi }\right) ,  \label{Eq.DispRel}
\end{equation}%
with the magnetic free energy (assuming a zero in-plane anisotropy ) $%
E=-HM_{s}\cos (\varphi -\varphi _{H})\sin \theta +\frac{H_{\mathrm{eff}}M_{s}%
}{2}\sin ^{2}\theta \sin ^{2}\varphi ,$ $(H_{\mathrm{eff}}=4\pi
M_{s}-2K_{\perp }/M_{s})$, $\gamma =g\mu _{\beta }/\hbar $ ($g\approx 2.2$
for Co$_{80}$Ir$_{20}$\cite{Zhang2021}). $\theta ,$ $\varphi $ and $\varphi
_{H}$ are the angles of the magnetization and external magnetic field
vectors, respectively. $4\pi M_{s}$ is the shape anisotropy field and $%
H_{\perp }=2K_{\perp }/M_{s}$ is a uniaxial anisotropy field that, depending
on the sign of $K_{\perp },$ can favor an easy axis normal to the film plane 
$(K_{\perp }>0)$ or an easy plane $(K_{\perp }<0).$ We assume that the
magnetic field varies in the $x-y$ plane so that for the out of plane
angular variation the film is placed in the $x-z$ plane and hence the
equilibrium polar angle for the magnetization vector is $\theta =\pi /2$. We
have neglected the effects of damping in Eq. (\ref{Eq.DispRel}). Within this
approximation we can evaluate the derivatives of Eq. (\ref{Eq.DispRel}) at
the equilibrium angles of the magnetization to arrive to the following
expression:%
\begin{equation}
\left( \frac{\omega }{\gamma }\right) ^{2}=\left[ H\cos (\varphi -\varphi
_{H})-H_{\mathrm{eff}}\sin ^{2}\varphi \right] \left[ H\cos (\varphi
-\varphi _{H})+H_{\mathrm{eff}}\cos 2\varphi \right] .  \label{Eq.wgPlane}
\end{equation}%
For the particular cases in which the equilibrium angle $\varphi $ can be
obtained analytically ($\varphi _{H}=0$ and $\varphi _{H}=\pi /2$) we obtain
the well known Kittel equations: 
\begin{eqnarray}
\frac{\omega }{\gamma } &=&\sqrt{H\left( H+H_{\mathrm{eff}}\right) }%
\;\;\varphi _{H}=0  \label{Eq.wg0-90} \\
\frac{\omega }{\gamma } &=&0\;\;\;\;\;\;\;\;\;\;\;\;\;\;\;\;\varphi _{H}=\pi
/2,\;\;H<H_{\mathrm{eff}}  \notag \\
\frac{\omega }{\gamma } &=&H-H_{\mathrm{eff}}\;\;\;\;\;\;\;\;\varphi
_{H}=\pi /2,\;\;H>H_{\mathrm{eff}}.  \notag
\end{eqnarray}%
For other values of $\varphi _{H}$ the equilibrium angles $\varphi $ must be
calculated numerically. In the sample CoIr-2 we have found $H_{r}(\varphi
_{H}=0)=750$ Oe and $H_{r}(\varphi _{H}=90%
{{}^\circ}%
)=15800$ Oe that can be inserted into Eq. \ref{Eq.wg0-90} to obtain $H_{%
\mathrm{eff}}=12630$ Oe and $g=2.199.$ This $g-$value is similar to that
reported in Ref. \cite{Zhang2021} an we will use it for the rest of the
samples in which it was not possible to determine $H_{r}(\varphi _{H}=90%
{{}^\circ}%
)$ due to limitations in the magnetic field or the low amplitude of a very
broad signal. $H_{\mathrm{eff}}$ was then determined in these samples using
only the first of Eqs. \ref{Eq.wg0-90} and the corresponding results are
shown in Fig. \ref{Fig-Heff}. We can observe that for samples CoIr-1 and 3
the value of $H_{\mathrm{eff}}$ is significantly larger than $4\pi M_{s}$,
indicating that there is an additional source of anisotropy favoring the
in-plane alignment of the magnetization. For films CoIr-2 and 4, on the
other hand, the estimated anisotropy field is quite similar to the shape
anisotropy which means that additional sources of anisotropy are comparative
small or mutually cancel out. From the expression of $H_{\mathrm{eff}}=4\pi
M_{s}-2K_{\perp }/M_{s}$ and the average value of $M_{s}=$ $910\pm 40$ emu/cm%
$^{3}$, obtained from VSM magnetometry, we can estimate the value of the
perpendicular anisotropy for the whole set of films (see Table \ref%
{Table-FMR}). We can observe that the perpendicular anisotropy constant for
CoIr-1/3 is $K_{\perp }=$ $-(3-4)$ Merg/cm$^{3}$, which is almost an order
of magnitude larger than $K_{\perp }=$ $-0.5$ Merg/cm$^{3}$ found in the
other two samples$.$ Even if considering the small difference in saturation
magnetization between samples CoIr-1/3 and CoIr-2/4 ($\left\langle M_{s}(300%
\text{ K})\right\rangle _{1,3}/\left\langle M_{s}(300\text{ K})\right\rangle
_{2,4}=0.945/0.885$) determined in the previous section, the estimated
values of $K_{\perp }$ are still much larger in samples CoIr-1/3 than in
CoIr-2/4. The smaller absolute values of $K_{\perp }$ in CoIr-2/4 could be
related to the Pt surface induced out of plane anisotropy, already observed%
\cite{Bhatti2025CoIr} in CoIr/Pt multilayers, which competes with the shape
and magnetocrystalline anisotropies that favor the in-plane alignment. 

\begin{figure}[tbh]
\includegraphics[ width=12cm]{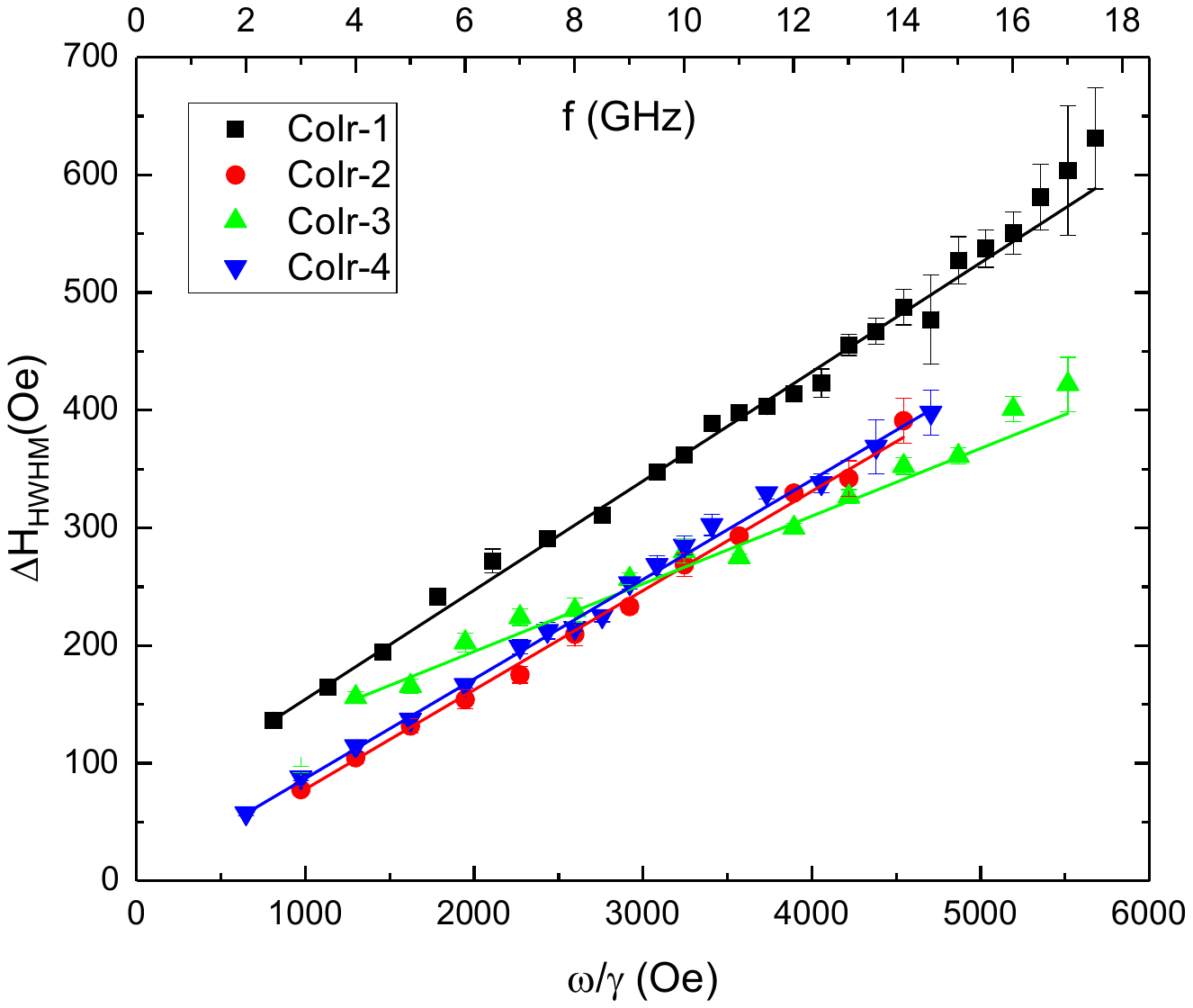}
\caption{Linewidth (half-maximum at half-height) as a function of frequency
obtained from broadband FMR measurements. Data have been fitted assuming the
linear dependence of Eq. \protect\ref{Eq.DH}.}
\label{Fig-DHrvsf}
\end{figure}

\begin{table}[h]
\caption{Parameters of all CoIr samples obtained from X-band resonator ($H_{%
\mathrm{eff}}$, $K_{\perp }$, $\Delta H_{\mathrm{pp}}$) and broadband ($%
\Delta H_{0}$, $\protect\alpha $) FMR measurements.}
\label{Table-FMR}\centering%
\begin{tabular}{cccccc}
sample & $H_{\mathrm{eff}}$ (kOe) & $K_{\perp }$ (Merg/cm$^{3}$) & $\Delta
H_{\mathrm{pp}}$(Oe) & $\Delta H_{0}$(Oe) & $\alpha $ \\ 
CoIr-1 & 18.2 & -3.06(15) & 407(5) & 77(8) & 0.088(2) \\ 
CoIr-2 & 12.6 & -0.54(15) & 295(5) & 6(3) & 0.084(2) \\ 
CoIr-3 & 20.3 & -4.05(15) & 313(5) & 80(8) & 0.058(2) \\ 
CoIr-4 & 12.6 & -0.52(15) & 300(5) & 2(3) & 0.085(2)%
\end{tabular}%
\end{table}
The largest absolute values of $K_{\perp }$ are of the order of the values
reported by different authors \cite%
{Kikuchi1999,Hashimoto2006,Nozawa2013,Zhang2014}, which have been always
attributed to the perpendicular [001] fiber texture present in the films and
the negative magnetocrystalline anisotropy of Co$_{80}$Ir$_{20}.$ However,
in our case there are different signs that point to the necessity of
considering strain effects. Microstructural studies presented in Section \ref%
{Section XRayDiffractometry} evidence that sample CoIr-1 has a wider
distribution of [001] axes than the rest of the films, which should reflect
in a smaller contribution to $H_{\mathrm{eff}}$ due to MCA. CoIr-2, 3 and 4
have almost the same width of the rocking curve, suggesting a similar
crystalline texture. CoIr-1 also has the largest FMR linewidth at X-band, as
can be seen in Table \ref{Table-FMR}. To get a deeper insight of the
microstructural influence on the magnetic anisotropy, we have performed
broadband (2 GHz - 18 GHz) FMR in all samples and analyzed the frequency
dependence of the resonance field and the linewidth. From the resonance
field we obtained similar $H_{\mathrm{eff}}\;$values than those found using
a resonator. For the analysis of the linewidth we assumed a frequency linear
dependence of the form\cite{Alvarez2013,Velazquez2021} 
\begin{equation}
\Delta H_{\mathrm{HWHM}}=\Delta H_{0}+\alpha \frac{\omega }{\gamma },
\label{Eq.DH}
\end{equation}%
where $\alpha $ is the damping parameter and $\Delta H_{0}$ is the extrinsic
contribution to the linewidth, related to the presence of inhomogeneities,
that tend to give a frequency independent contribution. The half width at
half maximum linewidth was obtained from the measured peak-to-peak linewidth
by assuming a lorentzian line shape ($\Delta H_{\mathrm{HWHM}}=\Delta H_{%
\mathrm{pp}}\sqrt{3}/2$). From the linear fit of the data presented in Fig. %
\ref{Fig-DHrvsf} we obtained the values $\Delta H_{0}$ and $\alpha $ shown
in Table \ref{Table-FMR}. We can see that the damping parameter is
approximately the same in all CoIr films in contact with Pt, $\alpha \approx
0.085,$ which is larger than those reported in Ref. \cite{Zhang2014} for Co$%
_{83}$Ir$_{17}$ ($\alpha \approx 0.07$) films with thicknesses in the range
55 nm $<t<$ 270 nm. Interestingly, they report that $\alpha $ increases from 
$\alpha =0.066$ for $t=270$ nm to $\alpha =0.071$ for $t=55$ nm which
suggests that larger $\alpha $ values could be expected for thinner films.
The sample CoIr-3, which is only in contact with Ta, shows a smaller damping
constant, $\alpha \approx 0.058.$The increase in $\alpha $ for CoIr films
that share an interface with Pt is related to the stronger interfacial
spin-mixing conductance $g^{\uparrow \downarrow }$ in a Pt/FM than in a
Ta/FM interface. For example, Pt/Py interfaces have\cite{You2021} $%
g^{\uparrow \downarrow }\sim 10$ nm$^{-2}$ while in Ta/Py\cite{Gomez2014} $%
g^{\uparrow \downarrow }\sim 1.3$ nm$^{-2}$. Larger values of $g^{\uparrow
\downarrow }$ cause a more efficient spin pumping injection from the
ferromagnet which allows a more efficient energy dissipation resulting in an
enhanced damping constant. The extrinsic linewidth contribution $\Delta H_{0}
$ is considerably larger in CoIr-1/3 than in CoIr-2/4 samples, which
correlates with structural and magnetic results already discussed. Note that
CoIr-1/3 have similar $\Delta H_{0}$ values and hence the larger $\Delta H_{%
\mathrm{pp}}$ in CoIr-1 compared to CoIr-3 found at X-band can be explained
by considering the different $\alpha $ values in both samples. 

Returning to the differences in the crystalline texture deduced from the
rocking curves, this would imply that sample CoIr-1 should have the lowest
magnetocrystalline contribution to the magnetic anisotropy and that CoIr-2,
3 and 4 should have similar anisotropy values, at variance with the observed
behavior. As we have found similar strain values in CoIr-1 and 3 and CoIr-2
and 4 we then analyze the potential contribution of magnetoelastic effects
on the anisotropy. We can relate the magnetoelastic contribution to the
anisotropy field with the magnetostriction and stress of the CoIr layer by
the expression\cite{Alvarez2015}:\ 
\begin{equation}
H_{\perp }^{ME}=\frac{2K_{\perp }^{ME}}{M_{s}}=-3\frac{\lambda _{xy}}{M_{s}}%
\sigma _{xy,}  \label{Eq.HME}
\end{equation}%
where $\lambda _{xy}$ is the magnetostriction coefficient in the hcp basal
plane and $\sigma _{xy}$ is the stress associated to the $\varepsilon _{33}$
strain,%
\begin{equation}
\sigma _{xy}=\left( c_{13}-\frac{c_{33}\left( c_{11}+c_{12}\right) }{2c_{13}}%
\right) \varepsilon _{33.}  \label{Eq.sxy}
\end{equation}%
This expression is obtained by considering a biaxial stress in a highly
textured sample. In order to evaluate the magnitude of $H_{\perp }^{ME}$ we
should have magnetostriction and elastic constant values for the Co$_{80}$Ir$%
_{20}$ alloy, which are not available. For an order of magnitude estimation
we could use the values reported\cite{Sander1999,Oliveira2026} for pure Co: $%
\lambda _{xy}\approx -80\times 10^{-6}$ and $c_{11}=307$, $c_{12}=165$, $%
c_{13}=107,$ $c_{33}=358$, all in GPa. Note that because we have a biaxial
strain we have used the average value between the two in-plane
magnetostriction constants, $\lambda _{A}\approx -50\times 10^{-6}$ and $%
\lambda _{B}\approx -107\times 10^{-6}.$ With these values and $M_{s}=$ $%
910\pm 40$ emu/cm$^{3}$ we can write for the anisotropy field $H_{\perp
}^{ME}[$Oe$]\approx 1.9\times 10^{6}\varepsilon _{33.}$ For strain values in
the range $\varepsilon _{33}$ $=-3\times 10^{-3}$, anisotropy fields of the
order of $\left\vert H_{\perp }^{ME}\right\vert \approx 6$ kOe can then be
reached by stress/strain effects. Note that Eq. \ref{Eq.HME} also gives the
sign of the anisotropy field which, using $\lambda _{xy}$ for pure Co, is
predicted to be positive and hence should favor an easy axis normal to the
film plane. However, because magnetocrystalline anisotropy and
magnetostriction are both related to spin-orbit coupling, a positive sign of 
$\lambda _{A}$ and $\lambda _{B}$ in Co$_{80}$Ir$_{20}$ is not unexpected
due to the different sign of $K_{MC}$. In any case, we have shown that
strain effects, which have been always overlooked in previous studies, must
be considered when evaluating the magnitude of magnetocrystalline anisotropy
in CoIr films. \bigskip \bigskip 

\section{Conclusions}

We have studied the influence of strain on the magnetic anisotropy of CoIr
films with negative magnetocrystalline anisotropy. We have found that the
characteristics of the underlayer strongly affects the strain of the
ferromagnetic layer which in turn leads to a magnetoelastic contribution to
the effective anisotropy field. The evaluation of $H_{\perp }^{ME}$ using
the magnetic parameters available for pure Co showed that magnetoelastic
effects could be of the same order than the MCA. The determination of the
elastic constants and, particularly, the sign and magnitude of the
magnetostriction constants of Co$_{80}$Ir$_{20}$ alloys are required for a
more accurate determination of strain effects. These results provide a
deeper understanding of how to tune the magnetic properties of CoIr films
through strain engineering, and their potential application in different
kinds of devices.

\bigskip

\section*{Acknowledgements}

Technical support from Rub\'{e}n E. Benavides, C\'{e}sar P\'{e}rez, and Mat%
\'{\i}as Guill\'{e}n is greatly acknowledged. This work was partially
supported by U.N. Cuyo 06/800202240100120UN, 06/80020240100273UN,
06/80020240100271UN, PIBAA 2022-2023 project MAGNETS Grant ID
28720210100099CO, ANPCyT PICT 2021-00113 project DISCO, all from Argentina.
We acknowledge the financial support of European Commission by the
H2020-MSCA RISE project ULTIMATE-I (Grant No. 101007825). Linguistic and
grammatical corrections were assisted by AI-based language models.

\bigskip

\bibliographystyle{unsrt}
\bibliography{Co80Ir20}
{}

\end{document}